\documentclass[prd,longbibiography,superscriptaddress,showpacs, showkeys,a4paper,notitlepage]{revtex4-1}
\usepackage{amssymb,amsmath,bm}
\usepackage{graphicx}

\newcommand{\Unitmat}{\mathbb{I}}
\newcommand{\Real}{\mathbb{R}}

\newcommand{\norm}[1]{\left\vert\left\vert #1 \right\vert\right\vert}
\begin{document}
\title{A Projective Interpretation of Some Doubly Special
 Relativity  Theories}
\author{N. Jafari}
\affiliation{Dept.\ of Physics, Semnan University, Semnan, Iran.}
\email{nosrat.jafari@gmail.com}  
\author{A. Shariati}
\affiliation{Physics Group, Faculty of Sciences, Alzahra 
University, Tehran, 19938, Iran.}
\email{shariati@mailaps.org}
\date{13 Aug 2011}
\published{15 Sep 2011}
\preprint{Phys. Rev. D 84, 065038 (15 Sep 2011)}
\begin{abstract}
A class of projective actions of the orthogonal group on the 
projective space is being studied.  It is shown that the 
Fock--Lorentz, and Magueijo--Smolin transformations known as
Doubly Special Relativity are such transformations. 
The formalism easily lead to new type transformations.
\end{abstract}
\maketitle

\section{Introduction}
The Lorentz symmetry is one of the cornerstones of modern physics.
It is the space-time symmetry of the Standard Model of Particle
Physics; and it is the symmetry of space-time as seen by
a freely falling observer, in a sufficiently small lab, in any
gravitational field.   In other words, it is respected both in
our quantum theories, and in our classical theories describing
gravity.  On the other hand, there are some arguments indicating
that at high enough energy scales, perhaps the Planck energy scale, 
Lorentz symmetry might be violated somehow.  
For example, in some approaches to quantum gravity
the spacetime has a polymer-like 
structure \cite{PhysRevD.59.124021}, in some the spacetime
is non-commutative \cite{PhysRevLett.87.141601}, and in some the 
spacetime has extra dimensions \cite{Burgess-et-al-02},
though there are
experimental and observational limits on such violations
\cite{JLM-Nature-03}.

To think about this problem, several approaches have been developed.
Some physicist have tried to consider the effect of various
Lorentz violating terms in the Lagrangian of Particle Physics (see
for example \cite{Colladay-Kostelecky-98, Coleman-Glashow99}).   
Some have tried to replace
a quantum deformation of the Poincar\'{e} group, a very important
example of which is the $\kappa$-Poincar\'{e} group 
\cite{Lukierski-et-al-91}.  And, some physcists have tried to
find a generalization of the Special Relativity.

One of the first generalizations of the Lorentz transformations
was introduced several decades ago by V.\ A. Fock \cite{Fock64},
whose motivation was to investigate the implications of the
relativity principle---equivalence of inertial frames---as far
as possible, that is, relaxing the constancy of the speed of light.
Later, S.\ N.\ Manida pointed out that these transformations
can be interpreted to exhibit a time-varying speed of 
light \cite{Manida99}.  Fock--Lorentz (FL) transformations are
transformations of the spacetime.  If one uses similar 
transformations for the energy-momentum space, one obtains
the so called Magueijo--Smolin (MS) transformations \cite{MS02}.

In Doubly Special Relativity (abbreviated DSR),
the idea is to find transformations which 
leave an energy (or length) scale invariant
\cite{Stepanov00, AC02a, MS02, MS03, KGN03, Kowalski-Glikman2005}.
Two famous examples of the DSR theories are the Amelino-Camelia 
\cite{AC02b, AC02a}, and the Magueijo-Smolin (MS) DSRs \cite{MS02}.
It is known that these DSRs are related to $\kappa$-Poincar\'{e}
(see \cite{Kowalski-Glikman2005}).
It is also possible to find more DSR theories from 
$\kappa$-Poincar\'{e} formalism \cite{LN03, Kowalski-Glikman2005}.

There has been some activities to understand the nature of the
nonlinear transformations of DSR theories.  To gain insight into these theories,
mathematical structures such as 
non-commutative geometry \cite{KGN03},
conformal groups \cite{Deriglazov04, Leiva05},
Finsler geometry \cite{Mignemi07},
five dimensional mechanics \cite{RD11},
and perhaps other structures are being invoked.

Few years ago, we argued that some DSRs,
namely the FL and the MS DSRs, are
merely re-descriptions of Einstein's Special Theory of Relativity \cite{JS04,jafari:462}, a view which some physicists do not agree (see for example \cite{AC11}).

Here we want to present a very simple geometrical interpretation
of the FL and the MS transformations, which will shed light onto 
these DSR theories.   This geometrical interpretation is so simple 
that one wonders why it has not been emphasized in the 
literature~\footnote{While this work was under review by the 
referees, we found that it is mentioned briefly 
in~\cite{Petkov2010}.}.
In spite of its simplicity, it enables one to find some new
transformations (see eqs. \ref{spacelike1}, \ref{spacelike2},
\ref{lightlike1}, and \ref{lightlike2}).  In this article
we  are restricting ourselves
to introduce this \textit{projective similarity} picture, 
postponing the study of its implications to a 
separate article.

It should be emphasized that we are not dealing with 
$\kappa$-deformed DSRs, which are deformations of the
Hopf algebra of the generators.

\section{The Real Projective Space}
We have to begin with a short review of the structure of the
Real Projective Space.  The subject is well known, and there are
several textbooks available (see for example \cite{Coxeter, Casse,  Liebscher}).  But a general knowledge of the basic definitions and properties is enough to follow the arguments.

The Real Projective Space of dimension $n$, denoted by
$\Real P^n$, is the space of \textit{rays} in $\Real^{n+1}$.
Any ray in $\Real^{n+1}$ is fully characterized by a pair of
antipodal points on the sphere 
$S_n = \left\{ x \in \Real^{n+1} \; \big\vert
\; \norm{x} := \sqrt{\sum_1^{n+1} x_i^2} = 1 \right\}$.
Therefore, $\Real P^n$ could be imagined as $S_n$ divided by the following equivalence relation.
\begin{equation} x, y \in S_n, 
\hskip 5mm plus 1mm minus 1mm x 
\sim y \quad \Leftrightarrow \quad
x = -y \; \vee \; x = y.\end{equation}
Usually this is written as 
$ \Real P^n = S_n / Z_2$, where $Z_2$ is the group consisting of
numbers $\left\{-1, 1 \right\}$.  The sphere $S_n$ is naturally 
endowed with a Riemannian metric, and its curvature is +1.
The above quotient map will induce a metric on $\Real P^n$.  So naturally $\Real P^n$ is a constant positive curvature space.

$\Real P^n$, could also be imagined as $\Real^n$ plus a ``(hyper-)
plane at infinity'' thus:
Divide the set of all lines of $\Real^n$ by the equivalence 
relation of parallelism. The resulting set would have one ``point'' 
for each class of parallel lines, which is called the``point at 
infinity'' of that class of parallel lines.  The totality of these 
points form a (hyper)plane, called the ``plane at infinity''.  Add 
this plane to the original $\Real^n$ and the result would be 
$\Real P^n$.  Now any class of parallel lines are ``meeting'' at a 
point at infinity, on the ``plane at infinity''.  We denote this 
plane at infinity with $L_\infty$.  When thinking of $\Real P^n$ in 
this way, one usually forgets that it is a curved space.

\section{Action of $SO(n)$ on $\Real P^n$}

The group $SO(n+1)$ acts naturally on $S_n$, and therefore on
$\Real P^n$.   In the following we will see that the group
$SO(n)$ also acts on $\Real P^n$.

The group $SO(n)$ acts naturally on the space $\Real^n$.  Let
$R$ denotes one such rotation.  The action of $R$ on $\Real^n$
could be extended to a map of $\Real P^n$ onto itself.  To this
end, we first remind that to each \textit{direction} in the $\Real^n$, there corresponds \textit{one} point at infinity.  Because of this,
the result of any rotation by $\theta = \pi$ would map any point
$P \in L_\infty$ into itself.  So that the action of the group
$SO(n)$ on $L_\infty$ would cover this line \textit{twice}, not once.
We say that $L_\infty$ is mapped onto itself under $SO(n)$, which means that it does not mix with any other plane.    In $\Real P^n$, and hence in $\Real^n$, there are no other planes which are being mapped to themselves under the action of \textit{all} members of $SO(n)$.

Now we are going to study another action of
$SO(n)$ on $\Real P^n$ which maps another plane, not necessarily the plane at infinity, onto itself.   For simplicity of argument, we
first present the case of $n=2$.

In $\Real^2$,
consider a line, not passing through the origin. We have excluded the origin which is the fixed point of $\Real^2$ under $SO(2)$.  

The most general line, in $(x,y)$ plane, not passing through the origin has the equation
\begin{equation} 1 + a\, x + b\, y  = 0, \qquad
(a, b) \neq (0, 0). \end{equation}
Let $\delta$ denotes this line, and let $H_\delta$ denotes the plane
$\Real^2$ minus this line, that is
\begin{equation} H_\delta = \left\{ (x,y) \in \Real^2 \; \big\vert
\; 1 + a\, x + b\, y \neq 0 \right\}. \end{equation}
Now consider the following transformation, from $(x,y)$ plane into
$(X,Y)$ plane:
\begin{equation} (x,y) \mathop{\longmapsto}^{f}
\left( X = \frac{x}{1 + a\, x + b\, y}, \;
Y = \frac{y}{1+ a\, x + b\, y}\right). \end{equation}
Obviously, this transformation is not defined on $\delta$.
It can be easily shown that the inverse of this transformation is
\begin{equation}  (X,Y) \mathop{\longmapsto}^{f^{-1}}
\left( x = \frac{X}{1 - a\, X - b\, Y}, \;
y = \frac{Y}{1 - a\, X - b\, Y}\right), \end{equation}
which is not defined on the line $\Delta$
\begin{equation} \Delta = \left\{ (X,Y)\; \big\vert \;
1 - a\, X - b\, Y = 0 \right\}, \end{equation}
it is defined on
\begin{equation} H_\Delta = \left\{ (X,Y) \in \Real^2 \; \big\vert
\quad 1 - a\, X - b\, Y \neq 0 \right\}. \end{equation}
Note that
\begin{equation} 1 - a\, X - b\, Y =
\frac{1}{1 + a\, x + b\, y}. \end{equation}
One can therefore interpret these mappings from $\Real P^2$ to itself, thus:

\begin{align} 
f(\delta) &= L_\infty, \qquad f(L_\infty) = \Delta, \\[0mm]
f^{-1}(L_\infty) &= \delta
, \hskip 10mm plus 1mm minus 1mm
f^{-1}(\Delta) = L_\infty
. \end{align}

Now, let $R$ be a rotation acting on the
$(X,Y)$ plane, 
\begin{equation}
R: \left( \begin{array}{l} X \cr Y \end{array} \right)
\mapsto \left( \begin{array}{l} X\, \cos\theta - Y\, \sin\theta
\cr X\, \sin\theta + Y\, \cos\theta \end{array} \right)
\end{equation}
and construct the following mapping of $\Real P^2$ onto itself.
\begin{equation}
S = f^{-1}\circ R \circ f.
\end{equation}
By straightforward calculation, one can easily show that the 
map $S:(x,y)\mapsto (x',y')$ is the following.
\begin{align}
x' &= \frac{x \, \cos\theta - y\, \sin\theta}{D}
\\[0mm]
y' &= \frac{ x\, \sin\theta + y\, \cos\theta}{D}, 
\end{align}
where
\begin{equation}
D = 1 + x (a - a\, \cos\theta - b\, \sin\theta)
+ y (b - b\, \cos\theta + a\, \sin\theta).
\end{equation}

\begin{figure}
\includegraphics[scale=0.6]{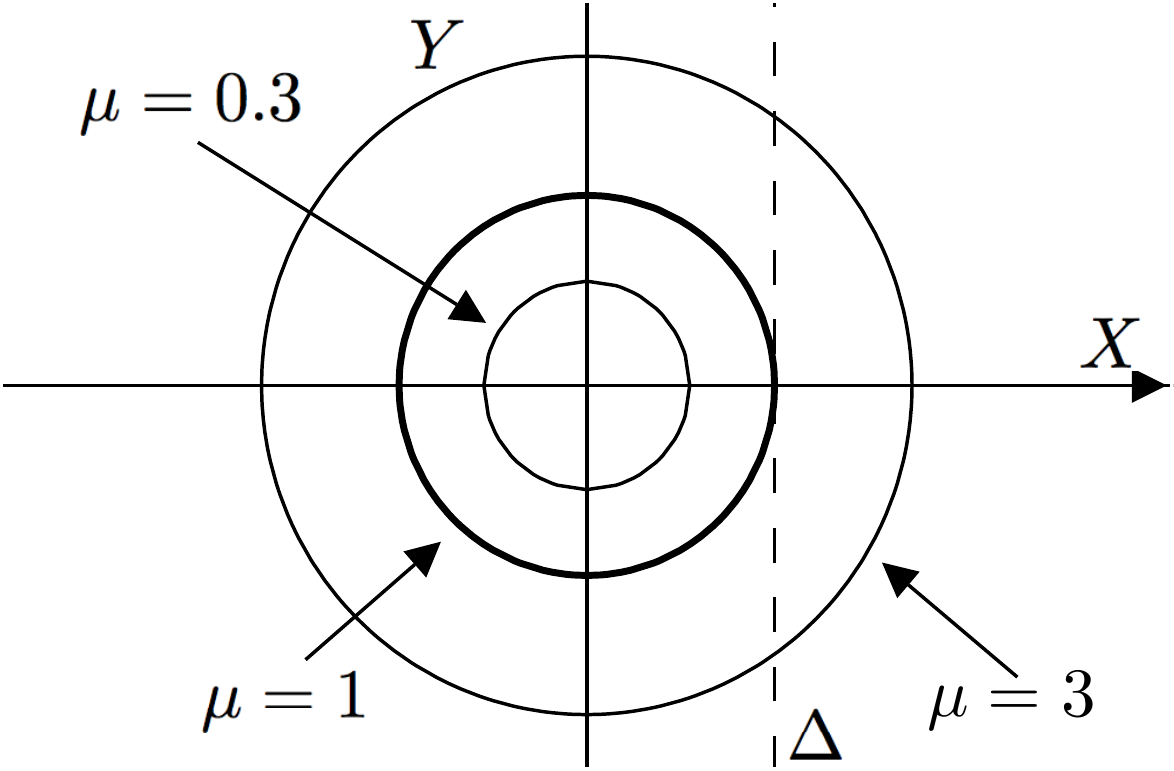}
\caption[]{The orbits, under the action of $SO(2)$, when $SO(2)$ is acting on $\Real P^2$ ordinarily, that is, when it is $L_\infty$ which is mapped into itself.  The line $\Delta$, with equation $X = 1$ is being mapped into $L_\infty$ by the projective map
$(X,Y) \mapsto \left( \frac{X}{1-X}, \frac{Y}{1-X}\right)$.
The touching circle is the circle with $\mu=1$.  Note that circles with $\mu >1$ cross $\Delta$ at two points.
}
\end{figure}

To have a clear picture in mind, let's take $a =1$ and 
$b = 0$~\footnote{By choosing $a=1$ we are choosing the 
$a^{-1}$ to be the unit of length.  This will simplify the formulas, and it is easy to change the unit length by simple dimensional 
arguments.}.   The denominator of the above transformations now
read $1 + x - x\, \cos\theta + y\, \sin\theta$, which for
$x = -1$ yield $\cos\theta + y\, \sin\theta$, from which it 
follows that
\begin{equation} (x = -1, y) \mathop{\longmapsto}^{S_\theta}
\left( x' = -1, y' = \frac{y - \tan\theta}{1 + y\, \tan\theta}
\right), \end{equation} 
which explicitly shows that under the above transformation
the line $\delta$ with equation $ 1 + x = 0$, is mapped into 
itself.  This line is not ``invariant'', in the 
sense that its points are not fixed.  Also note that 
$\theta$ and $\theta + \pi$ map $y$ to the same point $y'$, and
therefore, for $\theta \in \left[-\pi, \pi\right]$ the line
$x=-1$ is mapped to itself twice.

The line $\Delta$ is the line
 $X = 1$.   In the $(X,Y)$ plane, the orbits, under the action of $SO(2)$ are the circles with center at the origin, which are
characterized by a real non-negative number $\mu$.
\begin{equation} C_\mu := \left\{ (X,Y)\; \big\vert \;
X^2 + Y^2 = \mu \right\}. \end{equation}
Among these circles, there is one, and only one, which is tangent to the line $\Delta$.  We call it the \textit{touching} circle.
It is a straightforward calculation of elementary analytic geometry to show that for touching circle
\begin{equation} \mu = \frac{1}{a^2 + b^2}. \end{equation}
The corresponding sets, invariant under the action of $SO(2)$ on the $\Real P^2$, are the following conics.
\begin{equation} \Gamma_\mu := \left\{ (x,y) \; \big\vert \;
\frac{x^2 + y^2}{(1+x)^2} = \mu \right\}. \end{equation}
\begin{figure}
\includegraphics[scale=0.6]{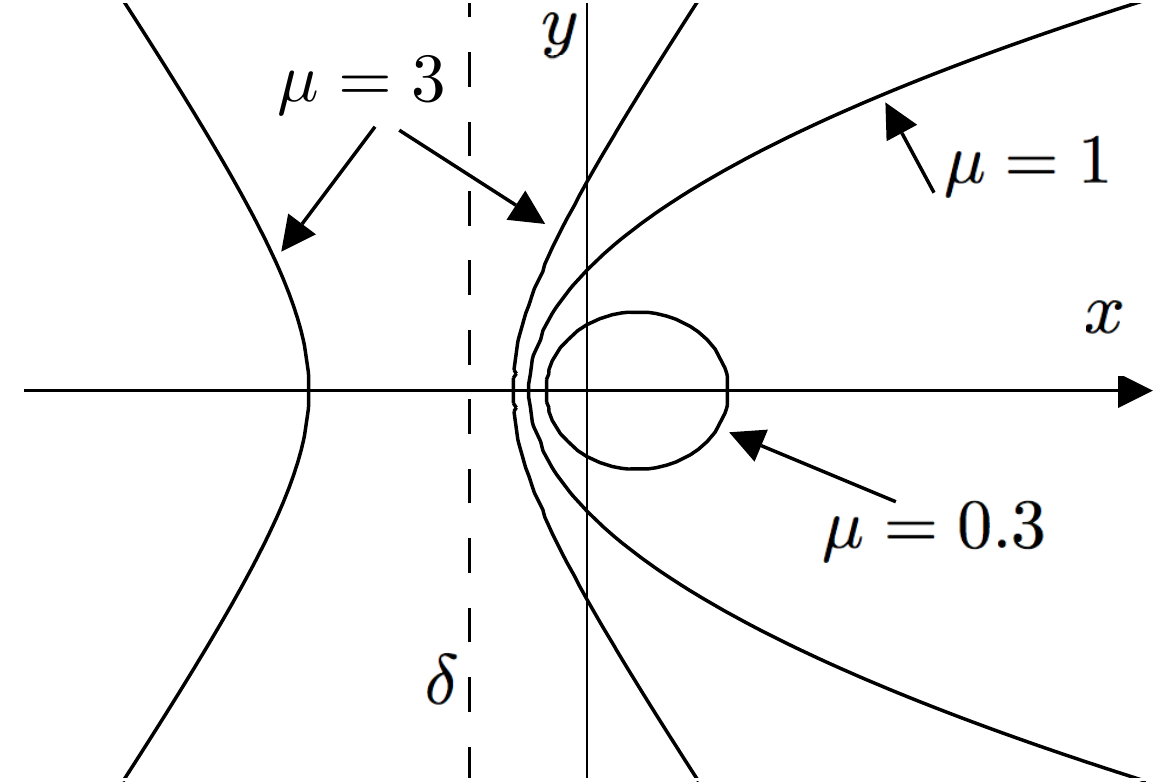}
\caption[]{The orbits, under the action of $SO(2)$, when 
$SO(2)$ is acting on $\Real P^2$ such that the line $\delta$, with
equation $x = -1$ is mapped into itself.  
The image of the touching circle, is now the parabola $\mu = 1$.
The line $\delta$ is invariant under this action.  As is seen from the two-sheet hyperbola, for some values of the rotation 
parameter $\theta$, points on one side of $\delta$ will move to 
the other side.   Also note that when $\mu \to \infty$, the two
sheets of the hyperbola will approach $\delta$, and will cover it twice.  
}
\end{figure}

Some remarks are worthy of mention.  First,
though the line $\delta$ cuts $\Real^2$ into two disjoint halves, the action of $SO(2)$ just constructed does not leave either of these halves invariant. 
This could be seen in Fig. 2.  Second, though the group $SO(2)$ 
is compact, the orbits $\Gamma_\mu$ for $\mu > 1$ are hyperbolas, which, are not compact.  The reason is that the map $f$ is singular
on $\delta$.

\section{The General Case}
To deal with $\Real P^n$, for $n>2$, and the more general case of $SO(1,n-1)$, let's use matrix notation. We define
\begin{equation} 
a = \left[\begin{array}{l} a^0 \cr a^1 \cr \vdots \cr a^{n-1} 
\end{array} \right], \qquad
 x = \left[ \begin{array}{l} x^0 \cr x^1 \cr
 \vdots \cr x^{n-1} \end{array} \right]. \end{equation}
and introduce the metric
\begin{equation} H = \left[\eta_{\mu\nu} \right] =
\mathop{\mbox{diag}}(-1, 1, \cdots, 1). \end{equation}
   The transformations $f$ and $f^{-1}$ have now a compact form---
\begin{equation} f: x \mapsto X = \frac{x}{1+a^T\, H\, x}, \end{equation}
\begin{equation} f^{-1}: X \mapsto x = \frac{X}{1-a^T\, H\, X},
\end{equation}
where a superscript $^T$ means matrix transposition.  
The above formula for $f^{-1}$ could be easily proved by substitution.

Now let's calculate the effect of $S = f^{-1}\circ \Lambda\circ f$ on $x$, where $\Lambda \in SO(1,n-1)$.  Under $f$ we have $X = \frac{x}{1 + a^T\, H\, x}$, and under
$R$ we have $X' = \frac{\Lambda\, x}{1+ a^T\, H\, x}$.  Finally, under
$f^{-1}$ we have $x' = \frac{X'}{1-a^T\, H\, X'}$.  So we need to calculate $1 - a^T\, H\, X'$ which is straightforward.
\begin{align} 
1 - a^T\, H\, X' 
& = 1 - a^T H \left( \frac{\Lambda\, x}{1 +
a^T\,H\, x} \right) 
\\[0mm] & = 1 - \frac{a^T\, H\, \Lambda\, x}{1 + a^T\, H\, x} 
\\[0mm] &= \frac{1 + a^T\, H\, x - a^T\, H\, \Lambda\, x}
{1 + a^T\, H\, x}. \end{align}
Therefore, we have
\begin{equation} \frac{1}{1 - a^T\, H\, X'} =
\frac{1 + a^T\, H\, x}{1 + a^T\, H \left( \Unitmat - \Lambda \right) x},
\end{equation}
where $\Unitmat$ is the unit $n\times n$ matrix.  
Now, for $S$ we have
\begin{align} 
x' & = \frac{X'}{1 - a^T\,H\, X'} 
\\[0mm] & =  
\frac{\Lambda\, H\, x}{1+ a^T\, H\, x} \cdot \frac{1 + a^T\, H\, x}
{1 + a^T\, H \left( \Unitmat - \Lambda \right) x}
\\[0mm] & = \frac{\Lambda\, x}{1 + a^T \, H \left( \Unitmat - 
\Lambda \right) x},
\end{align}
which, in the usual notation familiar in physics---with the Einstein summation convention understood---would read
\begin{equation} \label{trans} x'^\mu = \frac{\Lambda^{\mu}_{\;\nu}\, x^\nu}
{1 + a_\alpha \left( \delta^\alpha_{\;\beta} - \Lambda^\alpha_{\;\beta} \right) x^\beta}. 
\end{equation}

Perhaps the skeptical reader should check directly that the result of combining two such maps, first with $\Lambda$, and followed
with $\bar\Lambda$ would be one with $\bar\Lambda\, \Lambda$.
This is obvious from the construction of $S = f^{-1}\circ \Lambda 
\circ f$, but can also be checked by straightforward calculation.

The generators of the transformation (\ref{trans}) are
\begin{equation}
\tilde M_{\mu\nu} = x_\mu\, \partial_\nu - x_\nu\, \partial_\mu
+ \left( a_\mu\, x_\nu - a_\nu \, x_\mu \right) x^\sigma
\, \partial_\sigma.
\end{equation}
These operators satisfy the Lie algebra $so(1,n)$, because the
acting group is $SO(1,n)$.

\begin{figure}
\includegraphics[scale=0.6]{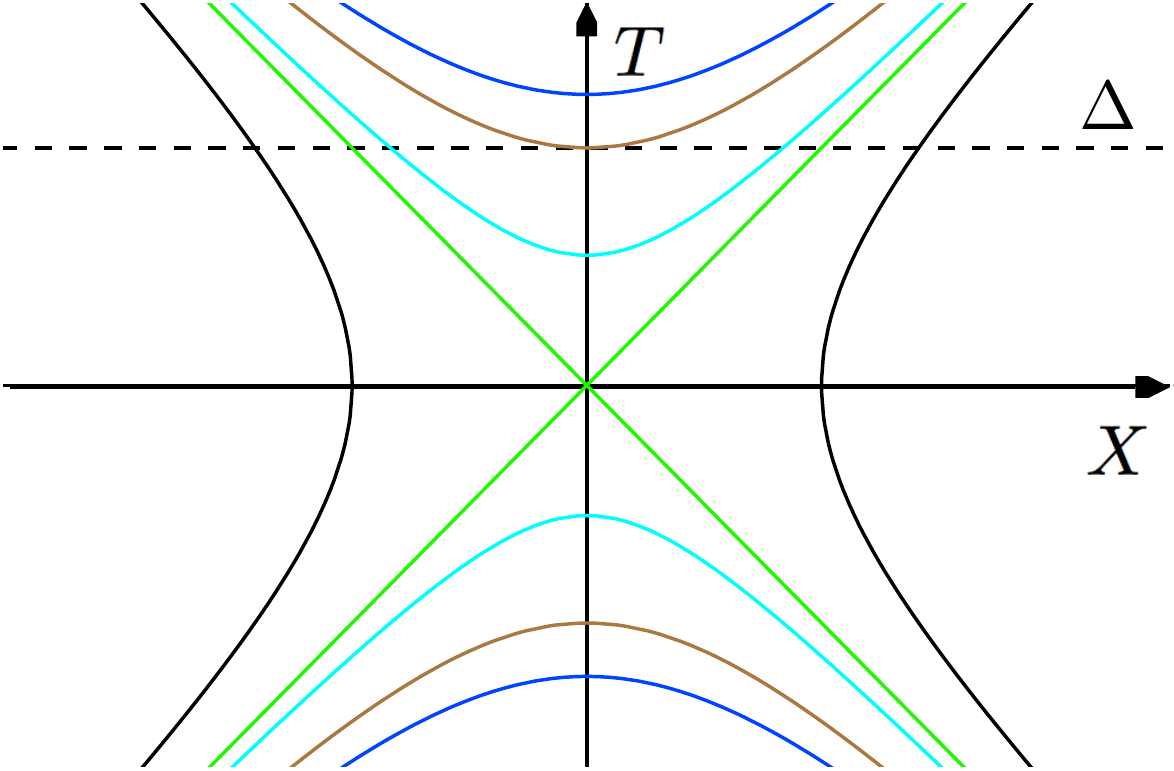}
\caption[]{Orbits of $SO(1,1)$ in $X^\mu$ space, for $a^\mu = (-1,0)$. The dashed line
with  equation $T = 1$ is the line $\Delta$.  The touching
hyperbola is the one tangent to the dashed line.  Note that all the orbits are asymptotic to the null directions.   }
\end{figure}

\section{The case $SO(1,1)$ acting on $\Real P^2$}
Under the action of $SO(1,1)$ the form
$-T^2 + X^2$ is invariant.  Now remind that
\begin{itemize} 
\item $-T^2 + X^2 = \rho < 0$ is a two-sheet hyperbola.
\item $-T^2 + X^2=0$ is the light cone, consisting of two straight lines passing through the origin.
\item $-T^2 + X^2 = \rho > 0$ is a two-sheet hyperbola.
\end{itemize}
First we must specify the vector $a$, which is either timelike ($a_\mu\, a^\mu < 0$), or lightlike ($a_\mu\, a^\mu = 0$), or
spacelike ($a_\mu\, a^\mu > 0$).

\subsection{A timelike $a$}
A simple form of such an $a^\mu$ is $a^\mu = (-1, 0)$.
The mappings $f$ and $f^{-1}$ now read
\begin{align}
 f&: x^\mu \mapsto X^\mu = \frac{x^\mu}{1 + t} \\[2mm]
 f^{-1}&: X^\mu \mapsto x^\mu = \frac{X^\mu}{1 - T}
\end{align}

\begin{figure}
\includegraphics[scale=0.6]{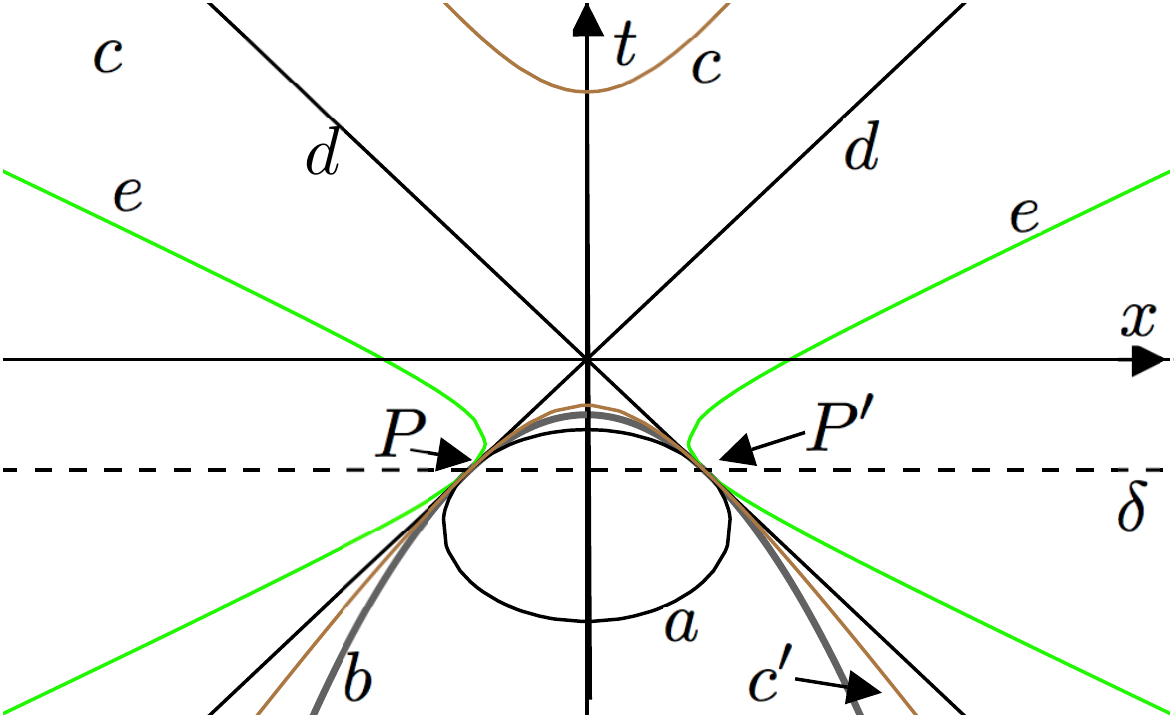}
\caption[]{Orbits of $SO(1,1)$ on $\Real P^2$, for 
$a^\mu = (-1,0)$. The invariant plane is $\delta$ 
(the set $t = -1$).
For $\mu < -1$ the orbit is an ellipse (labeled $a$).
For $\mu = -1$ the orbit is a parabola ($b$).
For $-1<\mu <0$ the orbit is a spacelike hyperbola of two sheets 
($c$ and $c'$). For $\mu = 0$ the orbit is a cone ($d$). 
For $\mu > 0$ the orbit is a time-like hyperbola of two 
sheets ($e$).  
Note that all these orbits intersect at the points
$P$ and $P'$, which are the images (under $f^{-1}$) of the null 
directions in the $X^\mu$ space, and remember that all the orbits 
of $SO(1,1)$ in $X^\mu$ space, are asymptotic to the null 
directions.   }
\label{fig4}
\end{figure}

The planes $\delta$ and $\Delta$ are the planes
\begin{align} & \delta: \qquad 1 + a_\mu\, x^\mu = 1 + t = 0, \\[1mm]
&\Delta: \qquad 1 - a_\mu \, X^\mu = 1 -  T = 0.
\end{align}
The touching hyperbola (in $X^\mu$ space) would be
the sheet $T>0$ of the hyperboloid
$ -T^2 + X^2 = -1 $, 
which by substitution  $ X^\mu = x^\mu /(1+t)$,
is mapped to the conic
$-t^2 + x^2 = - \left( 1 + t \right)^2$, 
which is the parabola 
$ t =  -\frac 1 2 \left( x^2  + 1\right)$.

 More generally, consider the conic
$ -t^2 + x^2 = \mu\left(1+t\right)^2$. 
For $\mu = -1$ this is the parabola mentioned above.  For 
$\mu \neq -1$ one can write it in the canonical form
\begin{equation} \frac{x^2}{1 + \mu} - 
\left( t + \frac{\mu}{1 + \mu} \right)^2 
= \frac{\mu}{\left( 1 + \mu \right)^2}.
\end{equation}
We therefore have
\begin{itemize}
\item for $-\infty < \mu < -1$, it is an ellipse.
As a submanifold of $\Real^2$ with the ordinary Euclidean 
topology, this set is compact.  This seems strange, since the
group $SO(1,1)$ is not compact, and we are saying that this 
ellipse is the image of $SO(1,1)$.   Here we should remember 
that the map $f$ is singular.
\item for $\mu = -1$, it is a parabola.
\item for $-1 < \mu < 0$, it is a space-like hyperbola, consisting 
of two sheets.
\item for $\mu = 0$, it is a 1 dimensional cone, consisting
of two lines.
\item for $0 < \mu < \infty$, it is a timelike hyperbola, 
consisting of two sheets.
\end{itemize}

Now let's look at the transformation (\ref{trans})
for the specific case of a boost in the $x^1$ direction, 
in which we have
\begin{equation} \left[\Lambda^\mu_{\;\nu} \right]
= \left[ \begin{array}{cc} 
\gamma & -\gamma\, v  \cr
-\gamma\, v & \gamma  \cr 
\end{array} \right], \qquad 
\gamma = \frac{1}{\sqrt{1-v^2}}. \end{equation}
For $a^0 = -1$, and $a^i = 0$, we have $a_0 = 1$, and $a_i=0$, and 
it follows that
\begin{equation} 1 + a_\alpha \left( \delta^\alpha_\beta -
\Lambda^\alpha_{\;\beta}\right) x^\beta 
= 1 - t \left( \gamma - 1 \right) + \gamma\, v\, x,
\end{equation}
and therefore the transformation $S = f^{-1} \circ \Lambda
\circ f$ reads
\begin{align}
t' &= \frac{\gamma \left( t -  v \, x\right)}
{1 - t \left( \gamma - 1 \right) + \gamma\, v\, x}, \\[1mm]
x' & = \frac{\gamma \left( x -  v\, t\right)}
{1 - t \left( \gamma - 1 \right) + \gamma\, v\, x}. 
\end{align}
This is the Fock-Lorentz transformation.
Using the same transformation on the energy-momentum space, one 
could easily get the Magueijo--Smolin transformation, the 
motivation
of which was to have transformations with an invariant energy
scale \cite{MS02}.

\begin{figure}
\includegraphics[scale=0.6]{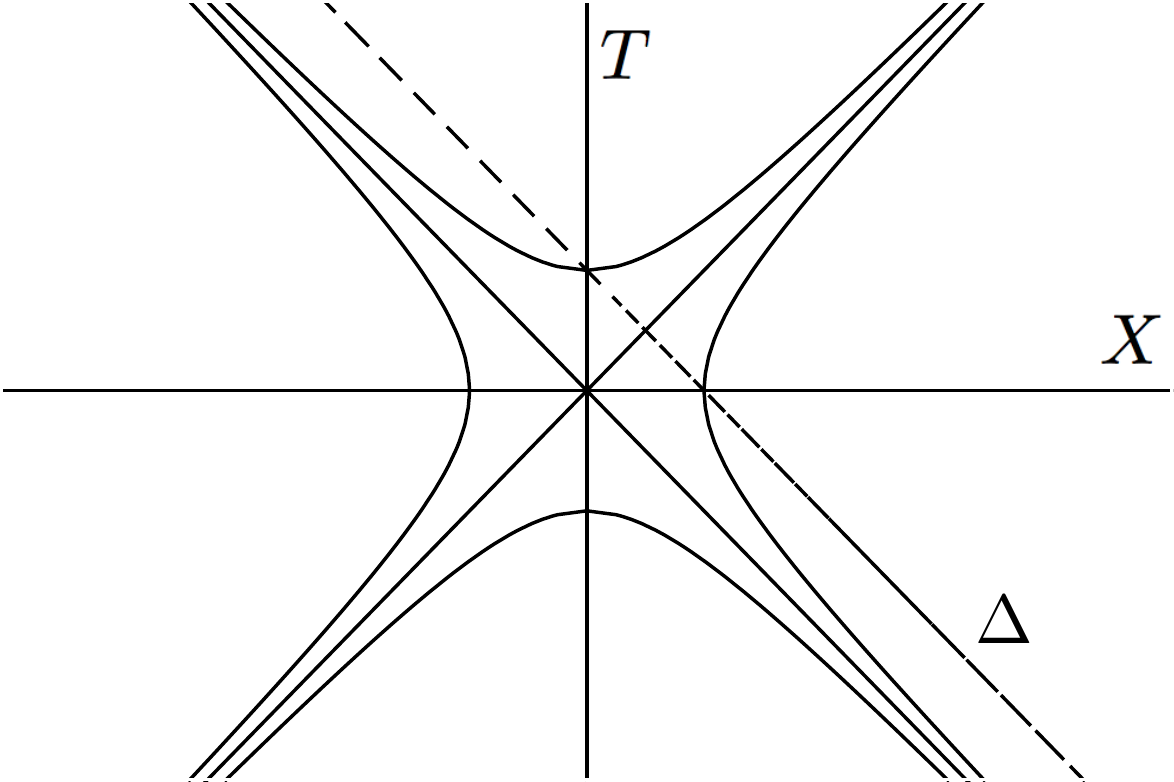}
\caption[]{The case $a^\mu = (-1,1)$.}
\label{fig5}
\end{figure}

\subsection{A spacelike $a$}
A simple form of such an $a^\mu$ is $a^\mu = (0, 1)$.  The mappings
$f$ and $f^{-1}$ now read
\begin{equation} X^\mu = \frac{x^\mu}{1 + x}, \qquad
x^\mu = \frac{X^\mu}{1-X}. \end{equation}
The planes $\delta$ and $\Delta$ are
\begin{equation} \delta: \quad 1 + x = 0, \qquad\qquad
\Delta: \quad 1 - X = 0. \end{equation}
The touching hyperbola (in $(T,X)$ plane) would be the sheet
$X>0$ of the hyperbola $-T^2 + X^2 = 1$.  The image of this 
hyperbola, under the map $f^{-1}$ is the parabola $x = -\frac 1 2
\left( t^2 + 1 \right)$.  It is obvious that all the other conics 
also can be derived from the previous case of a timelike $a$, by 
just a $\frac\pi 2$ rotation (Euclidean rotation) in the $(t,x)$ 
plane.  Now the ellipse of fig \ref{fig4} will be a 
a closed curve, turning back in time.  A physical interpretation 
of such a curve is very difficult.

Anyhow, let's find the ``projective'' transformation (\ref{trans}).
We now have
\begin{equation} 1 + a_\alpha \left( \delta^\alpha_\beta
-\Lambda^\alpha_{\;\beta}\right) x^\beta = 
1 - x\, (\gamma -1 ) + \gamma\, v\, t, 
\end{equation}
from which it follows that the boost 
in the $x$ direction, read
\begin{align} \label{spacelike1}
t' &= \frac{\gamma \left( t - v\, x \right)}
{1 - x (\gamma -1) + \gamma\, v\, t},
\\[1mm] \label{spacelike2}
x' & = \frac{\gamma \left( x - v\, t \right)}
{1 - x (\gamma -1) + \gamma\, v\, t}.
\end{align}

As far as we know, this transformation is new.

\begin{figure}
\includegraphics[scale=0.6]{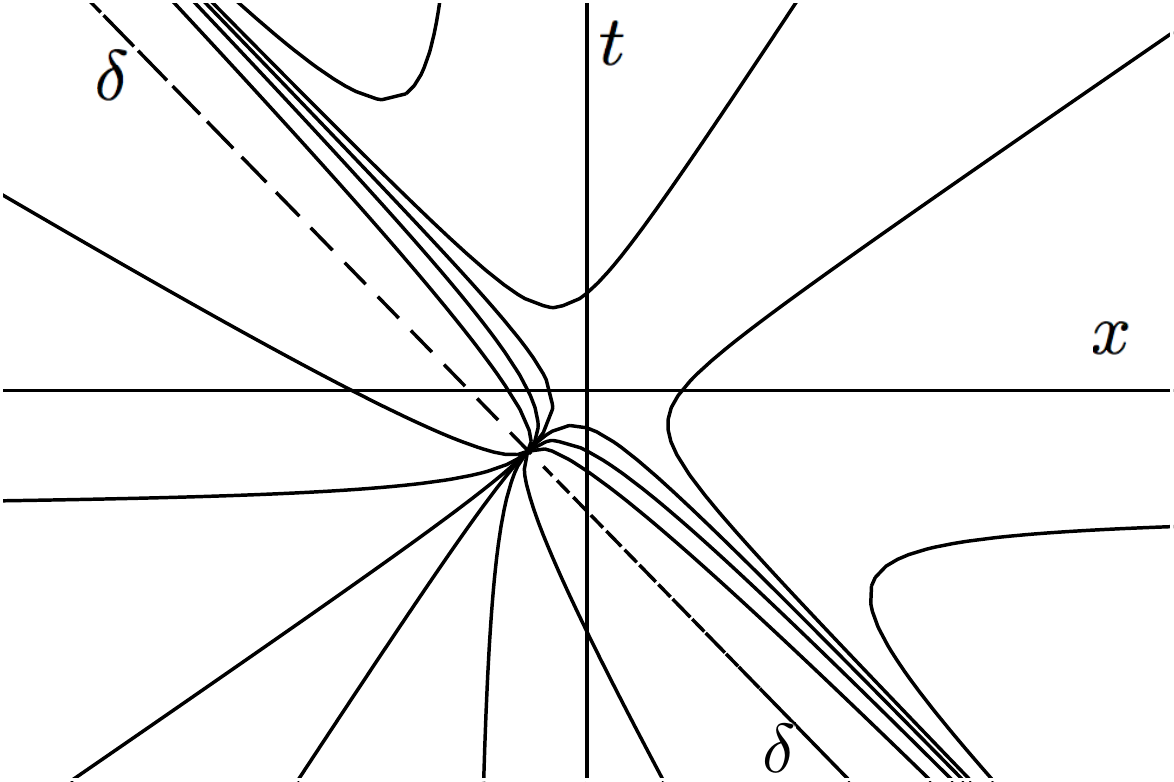}
\caption[]{The orbits of $SO(1,1)$ under the projective action for
$a^\mu = (-1,1)$.} 
\label{fig6}
\end{figure}

\subsection{A lightlike $a$}

Consider $a^\mu = (-1,1)$ which leads to 
$1 + a\cdot x = 1 + t + x$, and $1 - a \cdot X = 1 - T - X$.  
Now the line $\delta$ is $1+t+x=0$, and the line $\Delta$ is 
$1 - T - X =0$.  There is no touching hyperbola, since none of 
the hyperbolas are now tangent to $\Delta$.  The orbits of 
$SO(1,1)$ under this projective action are 
\begin{equation}
t^2 (\mu+1) + x^2 (\mu -1) + 2\, \mu \left( t + x + t \, x \right)
+ \mu = 0 \end{equation}
which are being drawn in fig \ref{fig6}.  We are not going to
describe these conics.   Let's only write the projective
transformation for this case.
\begin{equation}
1 + a_\alpha \left( \delta^\alpha_\beta - \Lambda^\alpha_{\;\beta}
\right) x^\beta = 1 + (t+x) \left(1-\gamma + \gamma\, v \right),
\end{equation}
from which it follows that
\begin{align} \label{lightlike1}
t' &= \frac{\gamma(t-v\, x)}
{1 + (t+x) \left(1-\gamma + \gamma\, v \right)},
\\[1mm] \label{lightlike2}
x' &= \frac{\gamma(x - v\, t)}
{1 + (t+x) \left(1-\gamma + \gamma\, v \right)}.
\end{align}
As far as we know, this transformation is also new.

\section{About the metric}
Using the map
\begin{equation}  
x \mapsto  X = \frac{x}{1 +  a \cdot x} 
\hskip 12mm plus 1mm minus 1mm
X^\mu = \frac{x^\mu}{1 + a_\alpha\, x^\alpha}
\end{equation}
one can pull back the flat metric 
$ds^2 = dX \cdot dX=\eta_{\mu\nu} \, dX^\mu \, dX^\nu $ to 
the space $H_\delta$.  The result would be
\begin{equation} ds^2 =
 \frac{dx \cdot dx}{\left( 1 + a \cdot x \right)^2}
- \frac{2\, a \cdot dx \;  x \cdot dx}
{\left( 1 + a \cdot x \right)^3} 
+ \frac{x \cdot x \; \left(a \cdot dx\right)^2 }
{(1 + a \cdot x)^4}.
 \end{equation}
This metric, being the pullback of the flat metric $\eta_{\mu\nu}$, 
has zero Gaussian curvature.  However, it must be noted that 
because of the Gauss-Bonnet theorem, $\Real P^2$ does not admit a 
global  Riemannian metric with zero Gaussian curvature---remind 
that the Euler characteristic of $\Real P^2$ is 1.  
Therefore, for $\Real P^2$, the pullback of the Euclidean metric 
could not be considered as a global metric on $\Real P^2$---it is 
not well defined on $\delta$---and the singularity is not removable 
by change of coordinates.  It seems to us that the case of the 
Minkowski metric on $\Real^4$, being pulled back to the 
corresponding $H_\delta$ has also this pathology; but at the moment 
we have no rigorous proof of that.

\section{Generalizing}
Generalizing to any matrix group acting on $\Real^n$,
for example to $SO(m,n-m)$, and in particular to 
$SO(1,n-1)$, is straightforward.  In the case of
$SO(1,n-1)$, three types of transformations will be obtained,
depending on the sign of $a \cdot a$.  The case of timelike 
$a^\mu = (-1, 0, 0, 0)$ 
will lead to the FL and MS transformations.  The only nontrivial
part of the generalization is the topology of the orbits,
which we are not going through in this paper.

\begin{acknowledgments}
 This work was partially supported by
 the Semnan University, 
 and partially by the
 research council of Alzahra University.
\end{acknowledgments}

\bibliography{JS2bibtex}

\end{document}